\documentclass[aps,prb,twocolumn,showpacs,showkeys,footinbib,superscriptaddress]{revtex4-2}%superscriptaddress

\usepackage{amsmath}
\usepackage{amssymb}
\usepackage{graphicx}
\usepackage{bm}
\usepackage{color}
\usepackage{relsize}
\usepackage{braket}
\usepackage[caption=false]{subfig}

\usepackage{hyperref}
\usepackage[all]{hypcap}

\usepackage{slashbox}
\usepackage{multirow}

\begin{document}
\title{Component exchange theory of trions }
\date{\today}

\author{Dinh Van Tuan}
\email[]{vdinh@ur.rochester.edu}
\affiliation{Department of Electrical and Computer Engineering, University of Rochester, Rochester, New York 14627, USA}
\author{Hanan~Dery}
\email[]{hanan.dery@rochester.edu}
\affiliation{Department of Electrical and Computer Engineering, University of Rochester, Rochester, New York 14627, USA}
\affiliation{Department of Physics and Astronomy, University of Rochester, Rochester, New York 14627, USA}

\begin{abstract} 
Treating the trion problem as an effective two-body system with exciton and electron components, we identify component exchange as the reason leading to trion formation. This mechanism can be visualized as a hole that toggles back and forth between two electrons.  The coined term component exchange is meant to inform that the repeated change of the electron component of the exciton originates from the exchange interaction. We develop a Bethe-Salpeter Equation for trions, showing that a bound trion state emerges if the electron and electron of the exciton are distinguishable particles (e.g., having opposite spins or residing in different valleys of the Brillouin zone). Similar to numerical techniques that treat the trion as a three-body problem, the trion Bethe-Salpeter Equation yields similar binding energy without using parameters beyond effective masses and dielectric constants.
\end{abstract}


\maketitle

%%%%%%%%%%%%%%%%%%%%%%%%%%%%%%%%%%%%%%%%%%%%%%%
%%%%%%%%%%%%%%%%%%%%%%%%%%%%%%%%%%%%%%%%%%%%%%%
%%%%%%%%%%%%%%%%%%%%%%%%%%%%%%%%%%%%%%%%%%%%%%%
%%%%%%%%%%%%%%%%%%%%%%%%%%%%%%%%%%%%%%%%%%%%%%%
\section{Introduction}

Much of the way we perceive excitons in semiconductors is attributed to the seminal work of Elliott, who tailored the hydrogen model to calculate exciton states  \cite{Elliott_PR57}. Using the electron and hole effective masses, the two-body problem successfully predicts the energy spectrum of charge-neutral excitons when employing appropriate Coulomb potential models to describe the dielectric screening \cite{Rytova_MSU67,Keldysh_JETP79,Cudazzo_PRB2011,Chernikov_PRL14,Stier_PRL18,VanTuan_PRB18,Meckbach_PRB18,Marauhn_PRB23}. Similar analogy can be drawn between a trion and hydrogen ion, H$^{-}$, where the trion complex is modeled as a three-body problem \cite{Stebe_PRL89,Kheng_PRL93,Finkelstein_PRL95,Astakhov_PRB00,Bracker_PRB05}. Solutions of this problem yield binding energies which agree with the measured energy difference between the trion and exciton resonances in the spectra of transition-metal dichalcogenides (TMD) monolayers \cite{Mayers_PRB15,Kylanpaa_PRB15,Kidd_PRB16,Donck_PRB17,Mostaani_PRB17,Filikhin_Nano18,VanTuan_PRL19}. An alternative approach to study the problem is to treat the trion as an exciton-electron system instead of a system with three independent bodies \cite{Shiau_PRB12,Sidler_NatPhys17,Efimkin_PRB17,Efimkin_PRB18,Fey_PRB20}, rendering the fact that the binding energy of an exciton far exceeds the binding energy between the exciton and electron. This approach renders the trion an effective two-body problem, comprising an exciton and a charge particle, which is compatible to the way trions are formed in optical experiments. The purpose of this work is to establish such equivalent exciton-electron picture without using phenomenological parameters or parameters beyond those used in the three-body problem (effective masses and dielectric constants). Hereafter, we focus on an exciton-electron system to make the discussion intelligible, and readers should bear in mind that an exciton-hole system is equivalent.

As we show in this work, an exciton-electron system becomes a bound trion if the hole can toggle back and forth between two distinguishable electrons. The repeated component exchange intertwines the exciton and electron, such that the resulting system is not that of an electron and rigid exciton. We construct an effective Bethe-Salpeter Equation to describe the exciton-electron system and then solve the equation straightforwardly by matrix inversion. Very good agreement is found between the results of this theory and solutions of three-body problems. In what follows we present the theory in Sec.~\ref{sec:the}, followed by discussion of the results in Sec.~\ref{sec:res} and conclusions in Sec.~\ref{sec:sum}. The Appendices include derivations and computational details.

%%%%%%%%%%%%%%%%%%%%%%%%%%%%%%%%%%%%%%%%%%%%%%%
%%%%%%%%%%%%%%%%%%%%%%%%%%%%%%%%%%%%%%%%%%%%%%%
%%%%%%%%%%%%%%%%%%%%%%%%%%%%%%%%%%%%%%%%%%%%%%%
%%%%%%%%%%%%%%%%%%%%%%%%%%%%%%%%%%%%%%%%%%%%%%% 
%%%%%%%%%%%%%%%%%%%%%%%%%%%%%%%%%%%%%%%%%%%%%%%
%%%%%%%%%%%%%%%%%%%%%%%%%%%%%%%%%%%%%%%%%%%%%%%
%%%%%%%%%%%%%%%%%%%%%%%%%%%%%%%%%%%%%%%%%%%%%%%
%%%%%%%%%%%%%%%%%%%%%%%%%%%%%%%%%%%%%%%%%%%%%%%

%%%%%%%%%%%%%%%%%%%%%%%%%%%%%%%%%%%%%%%%%%%%%%%%%%%%%%%%%%

\section{Theory}\label{sec:the}
The state of a noninteracting exciton-electron system is denoted by $| \mathbf{k}, \mathbf{Q}-\mathbf{k}\rangle$ where $\mathbf{k}$ is the exciton's center-of-mass wavevector and  $\mathbf{Q}-\mathbf{k}$ is that of the electron, written such that $\mathbf{Q}$ is the total wavevector of the system. The mass of the exciton is $m_x = m_e + m_h$ where $m_e$ and  $m_h$ are effective masses of its hole and electron, respectively, and $m_e'$ is the effective mass of the other electron in the system. In a semiconductor crystal, we say that the two electrons are indistinguishable if the electron of the exciton and the other electron have the same spin and they reside in the same valley of the same energy band. We identify this case in what follows by the parameter $\delta_d = 0$. Otherwise, the two electrons are distinguishable ($\delta_d = 1$).  Denoting the Coulomb interaction between the electron and the exciton’s electron (hole) by $V_{\text{ee}}$ ($V_{\text{eh}}$), the exciton-electron interaction is written as  
\begin{eqnarray}
M(\mathbf{Q},\mathbf{k},\mathbf{p}) &=& \langle \mathbf{p}, \mathbf{Q} -\mathbf{p} | V_{\text{ee}} + V_{\text{eh}} | \mathbf{k}, \mathbf{Q} - \mathbf{k}\rangle  \nonumber \\ 
&=&  D(\mathbf{k},\mathbf{p}) + X(\mathbf{Q},\mathbf{k},\mathbf{p}).\,\,\,
 \label{eq:M}
\end{eqnarray}

\begin{figure}[t] 
\centering
\includegraphics[width=8.5cm]{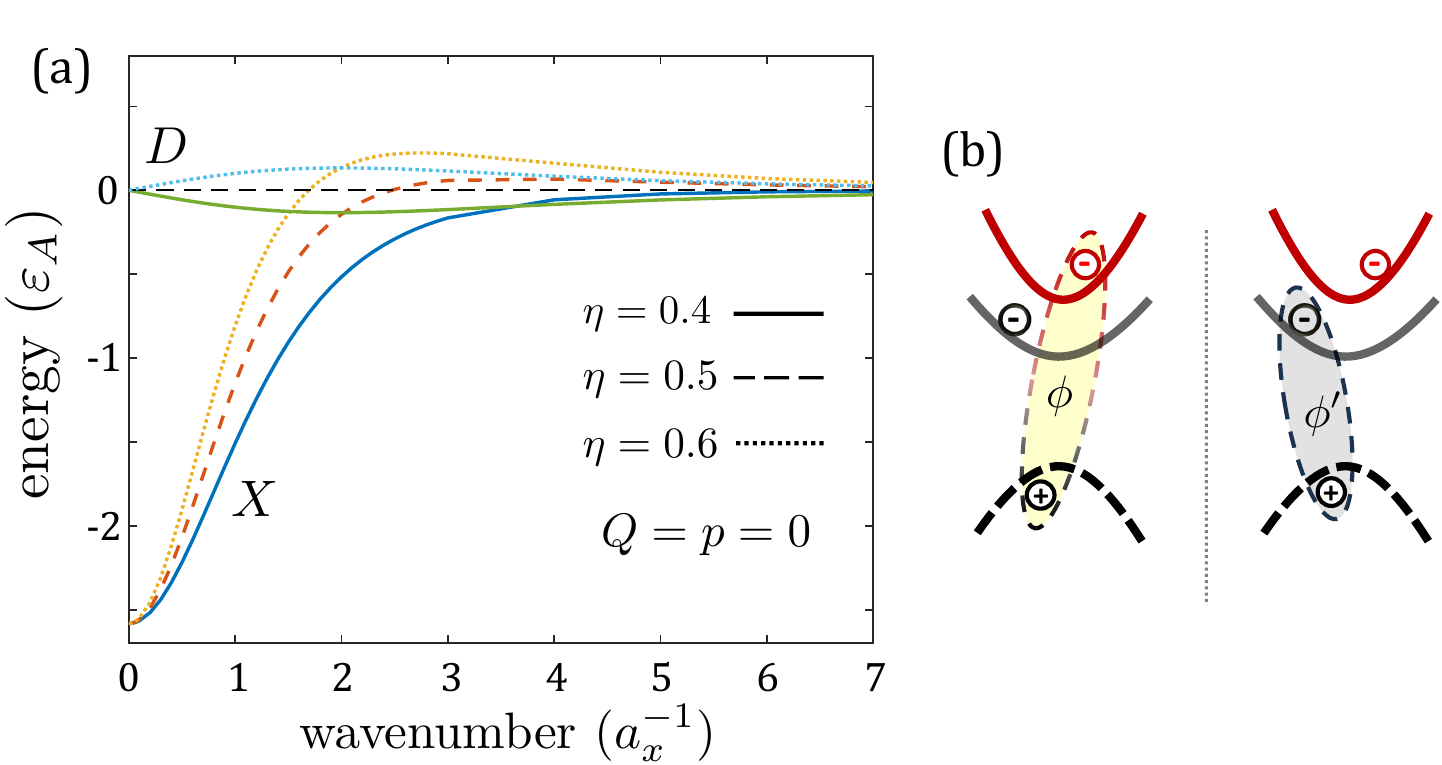}
\caption{ (a) Direct (D) and exchange (X) contributions to the interaction between an electron and exciton in the 1$s$ state of the 2D hydrogen model. The electron and the electron of the exciton are distinguishable. When the electrons are indistinguishable, the direct interaction remains the same and the exchange one becomes repulsive (switches sign). (b) Two views of an exciton-electron system with distinguishable electrons: one with exciton $\phi$ and electron $e'$ (bottom valley) and the other with exciton $\phi'$ and electron $e$ (top valley).}\label{fig:M} 
\end{figure}

The interaction is partitioned to direct and exchange contributions \cite{Ramon_PRB03,Shahnazaryan_PRB17,Yang_PRB22}, where the direct interaction is the part that is independent of electrons distinguishability (Appendix~ \ref{app:M}),
\begin{eqnarray}
D(\mathbf{k},\mathbf{p}) = \frac{V_{\mathbf{k}-\mathbf{p}}}{A} \sum_{\mathbf{q}} \left[  \phi_{\bar{\eta}\mathbf{p}+\mathbf{q}} \phi^\ast_{\bar{\eta}\mathbf{k}+\mathbf{q}} -   \phi_{\eta\mathbf{p}+\mathbf{q}} \phi^\ast_{\eta\mathbf{k}+\mathbf{q}} \right]\!.\,\,\,\,\,\,
 \label{eq:D}
\end{eqnarray}
$V_{\mathbf{k}}$ and $\phi_{\mathbf{k}}$ are Fourier transforms of the Coulomb potential and exciton wavefunction, respectively, and we have assumed a two-dimensional (2D) semiconductor system whose area is $A$. The other parameters are $\eta = m_e / (m_e + m_h)$ and $\bar{\eta}= \eta - 1$. The direct interaction in exciton-electron systems treats the exciton as a rigid body that keeps the same electron before and after the interaction. It can be readily seen that $D(\mathbf{k},\mathbf{p})=0$ when the effective masses of the electron and hole in the exciton are the same, $m_e= m_h$ (i.e., $\eta= -\bar{\eta} = 1/2$). The  direct interaction vanishes completely because the electron-electron and electron-hole interactions offset each other perfectly in this case. In addition, charge neutrality of the exciton means that the electron-electron and electron-hole interactions cancel each other in the long-wavelength limit regardless of the value of $\eta$ (i.e., $D(\mathbf{k} \rightarrow \mathbf{p}) = 0$). In terms of Eq.~(\ref{eq:D}), the square-brackets expression converges to zero faster than the Coulomb potential  divergence $V_{\mathbf{k}-\mathbf{p}} \propto 1/|\mathbf{k}-\mathbf{p}|$ when  $\mathbf{k} \rightarrow \mathbf{p}$. To assess the amplitude of the direct interaction, we assume an interaction between an electron and 1$s$ exciton state of a 2D-hydrogen model, $\phi_{\mathbf{k}}= \sqrt{8\pi}a_x/[1 + (ka_x)^2]^{3/2}$ where $a_x$ is the exciton's Bohr radius. The  lines denoted by $D$ in Fig.~(\ref{fig:M})(a) evaluate the direct interaction $D(\mathbf{k},\mathbf{p})$ when $p=0$ for $\eta=0.4$, 0.5 and 0.6. The energy units of the $y$-axis are $\varepsilon_{A} = 4a_xe^2/A\epsilon_r$ where $\epsilon_r$ is the dielectric constant of the environment surrounding the 2D system.

The exchange term in Eq.~(\ref{eq:M}) depends on whether the two electrons of the system are indistinguishable ($\delta_d = 0$) or distinguishable ($\delta_d = 1$). Following the derivation in Appendix~\ref{app:M},
\begin{eqnarray}
X(\mathbf{Q}, \mathbf{k},\mathbf{p}) &=& \frac{(-1)^{\delta_d}}{A} \sum_{\mathbf{q}} V_{\mathbf{q}} \, \phi'_{\eta'\mathbf{p}+\mathbf{k} - \mathbf{Q}- \mathbf{q}}  \nonumber \\
&& \times \left[ \phi^\ast_{\eta\mathbf{k}+\mathbf{p} - \mathbf{Q}}  -  \phi^\ast_{\eta\mathbf{k}+\mathbf{p} - \mathbf{Q}- \mathbf{q}} \right] .\,\,\,\,\,\,
 \label{eq:X}
\end{eqnarray}
Quantities with prime symbols mean that the exciton state and its mass ratio, $\phi'_{\mathbf{k}}$ and $\eta'= m_e' / (m_e' + m_h)$, are calculated when the  hole is paired with the other electron in the final state. Using the diagrams in Fig.~(\ref{fig:M})(b) as a reference, the exchange (direct) interaction corresponds to the transition $\phi \rightarrow \phi'$ ($\phi \rightarrow \phi$). The offset between the electron-electron and electron-hole interactions is less effective when the electron component of the exciton is exchanged. The effect is seen in the second line of Eq.~(\ref{eq:X}), where the argument of the first (second) wavefunction lacks (includes) the integration variable $\mathbf{q}$. These terms originate from the electron-hole (electron-electron) interaction, and they are least effective in cancelling each other in the long wavelength limit  $|\eta\mathbf{k}+\mathbf{p} - \mathbf{Q}| \ll a_x^{-1}$.

The lines denoted by $X$ in Fig.~(\ref{fig:M})(a) evaluate the exchange interaction $X(\mathbf{Q},\mathbf{k},\mathbf{p})$ when $Q=p=0$ for $\eta=\eta'=\{0.4, 0.5, 0.6\}$.  The results show the dominant contribution from exchange interaction compared with the direct one. This dominance is a universal property, not limited to 2D dimensional systems. Additionally, the behavior is qualitatively similar with Coulomb potentials that use dielectric screening models other than the ideal 2D case, $V_{\mathbf{q}} = 2\pi e^2/A\epsilon_r q$. This universality stems from charge neutrality of the exciton, where the direct interaction remains weak because the electron-electron interaction offsets the electron-hole interaction effectively if the hole is attached to the same electron before and after the interaction. The exchange interaction is not subjected to a similar strict constraint, rendering the interaction attractive (repulsive) when the electrons are distinguishable (indistinguishable) in the long-wavelength limit. This difference is expressed through the dependence of $X(\mathbf{Q},\mathbf{k},\mathbf{p})$ on $\delta_d$, originating from the  fermions anticommutation relations (Appendix~\ref{app:M}). As we show next, the outcome is that a trion state is formed when the two electrons are distinguishable.

%%%%%%%%%%%%%%%%%%%%%%%%%%%%%%%%%%%%%%%%%%%%%%%%%%%%%%%%%%

%%%%%%%%%%%%%%%%%%%%%%%%%%%%%%%%%%%%%%%%%%%%%%%%%%%%%%%%%%
\subsection{Trion Bethe-Salpeter Equation}
\begin{figure}
\includegraphics[width=7cm]{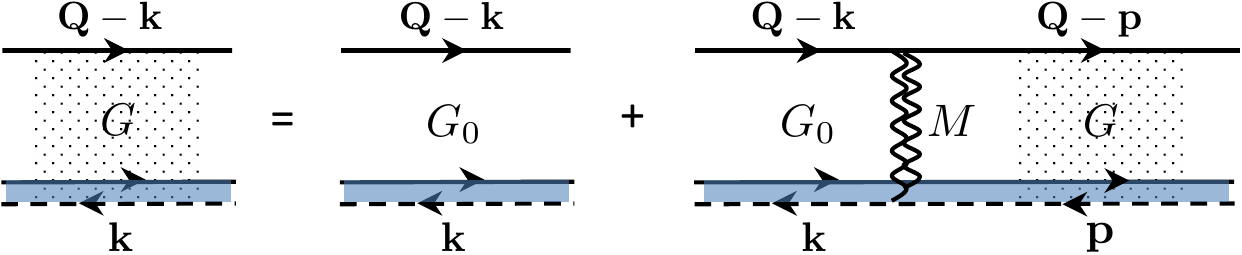}
\caption{Feynman diagram representation of the Bethe-Salpeter Equation of a trion with two indistinguishable electrons. The top (bottom) parts are electron (exciton) propagators where dashed/solid lines denote hole/electron components. The exciton-electron interaction (vertical wiggly lines), $M=D(\mathbf{k},\mathbf{p})+ X(\mathbf{Q}, \mathbf{k},\mathbf{p})$, is the sum of direct and exchange contributions in Eqs.~(\ref{eq:D})-(\ref{eq:X}) with $\delta_{d}=0$.} \label{fig:BSEid} 
\end{figure}

The Bethe-Salpeter Equation (BSE) describes bound states of two-body problems, where here these two bodies are the exciton and electron. Having established the exciton-electron interaction, the trion BSE of the indistinguishable case reads (Appendix~\ref{app:BSE})
 \begin{eqnarray}
G(\mathbf{k})  &=&  G_0(\mathbf{k}) + G_0(\mathbf{k}) \sum_{\mathbf{p}}M(\mathbf{Q}, \mathbf{k},\mathbf{p})G(\mathbf{p}).
 \label{eq:BSEid}
\end{eqnarray}
$G$ and $G_0$ are the interacting and noninteracting Green's functions of the exciton-electron system, respectively. Their dependence on conserved quantities such as energy (E) and total wavevector ($\mathbf{Q}$) is implied. The interaction $M(\mathbf{Q}, \mathbf{k},\mathbf{p})$ is given by Eqs.~(\ref{eq:M})-(\ref{eq:X}) with $\delta_d=0$. Equation~(\ref{eq:BSEid}) describes repeated interaction of the exciton and electron, represented by the ladder-type Feynman diagram in Fig.~\ref{fig:BSEid}. Practically, it means that we work in the dilute limit (i.e., no Fermi-sea of electrons nor a Bose-sea of excitons around the exciton-electron system), which in turn also means that we can neglect dynamical renormalization effects of the direct and exchange interactions. The noninteracting Green's function reads
 \begin{eqnarray}
G_0(\mathbf{k})  &=&  \frac{1}{E - \frac{\hbar^2 (\mathbf{Q}-\mathbf{k})^2}{2m_e'}  - \frac{\hbar^2 k^2}{2m_x} + i\delta},
 \label{eq:G0id}
\end{eqnarray}
where $\delta$ is a small positive energy that describes broadening effects. Note that $m_e' = m_e$ because the electrons are indistinguishable and that the energy $E$ is measured with respect to the exciton resonance energy (reference level in this case). If the solution of Eq.~(\ref{eq:BSEid}) yields a bound trion state, we recognize this state as a resonance in negative energy of the spectral function,
 \begin{eqnarray}
A_{\text{id}}(E,\mathbf{Q})  &\propto& -\mathcal{I}m \left \{ \sum_\mathbf{k} G(\mathbf{k}) \right \}.
 \label{eq:Aid}
\end{eqnarray}
As we will later show, the spectral function shows no bound state when the two electrons are indistinguishable.

%%%%%%%%%%%%%%%%%%%%%%%%%%%%%%%%%%%%%%%%%%%%%%%%%%%%%%%%%%
%%%%%%%%%%%%%%%%%%%%%%%%%%%%%%%%%%%%%%%%%%%%%%%%%%%%%%%%%%

\begin{figure*}
\centering
\includegraphics[width=16cm]{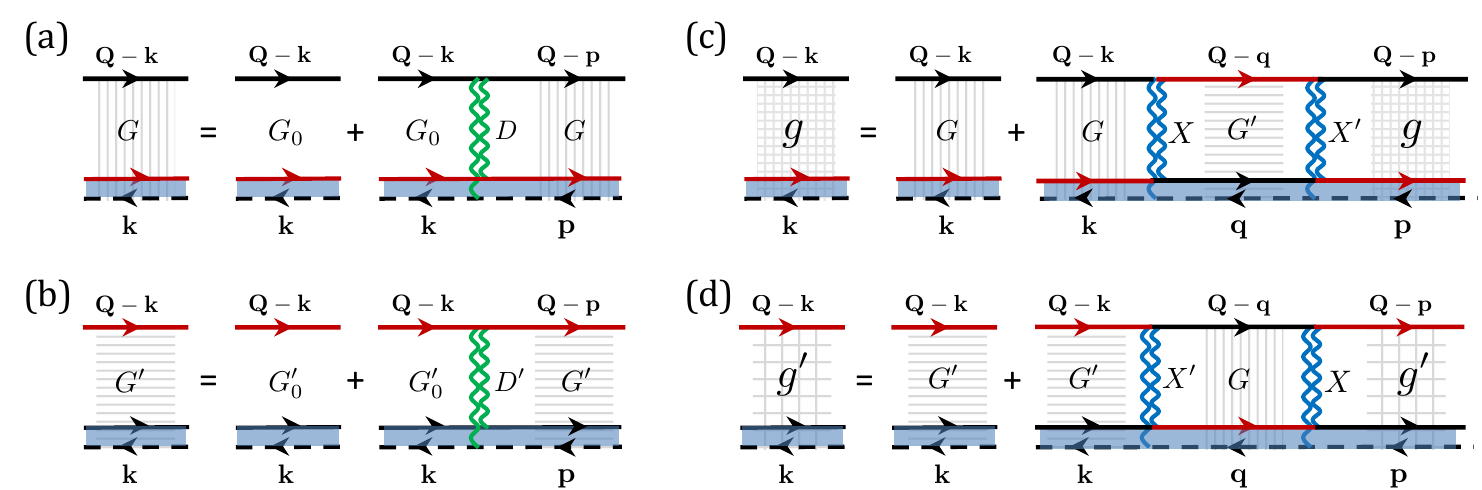}
\caption{ Feynman diagrams representation of the trion Bethe-Salpeter Equation with two distinguishable electrons. (a)-(b) The first step corresponds to dressing the noninteracting propagators by the direct interaction $D$, as given by Eq.~(\ref{eq:D}). (c)-(d) In the second step, the propagators are dressed by the exchange interaction, as given by Eq.~(\ref{eq:X}) with $\delta_{d}=1$. The use of two exchange interactions is required to bring the exciton back to its original state, where X followed by X' corresponds to the sequence $\phi \rightarrow \phi'  \rightarrow \phi$, and X' followed by X to $\phi' \rightarrow \phi  \rightarrow \phi'$.}\label{fig:BSEd} 
\end{figure*} 

Repeating the analysis for the distinguishable case is subtle because the exciton state may be different after each interaction depending on whether the direct or exchange interactions took place. To overcome this difficulty, we introduce a coupled set of hierarchical equations, $G_0 \rightarrow G \rightarrow g$, where we first calculate $G$ from $G_0$ by using the direct interaction, followed by dressing of $G$ by the exchange interaction to get the final Green's function, $g$. This hierarchical scheme renders the fact that two exchange interactions are needed to bring an exciton back to its original state and that the exciton-electron system can be dressed by direct interactions between two exchange interactions. The result of this hierarchy is shown by the Feynman diagrams in Fig.~\ref{fig:BSEd}. The first step corresponds to the diagrams in Fig.~\ref{fig:BSEd}(a)-(b), where as discussed before and visualized in Fig.~\ref{fig:M}(b), quantities with prime symbols mean that the hole is paired with the other electron. We get 
\begin{eqnarray}
G(\mathbf{k})  &=&  G_0(\mathbf{k}) + G_0(\mathbf{k})   \sum_{\mathbf{p}}D(\mathbf{k},\mathbf{p})G(\mathbf{p}), \nonumber \\
G'(\mathbf{k})  &=&  G'_0(\mathbf{k}) + G'_0(\mathbf{k}) \sum_{\mathbf{p}}D'(\mathbf{k},\mathbf{p})G'(\mathbf{p}),
 \label{eq:BSEdir}
\end{eqnarray}
with the noninteracting Green's functions
 \begin{eqnarray}
G_0(\mathbf{k})  &=&  \frac{1}{E - \frac{\hbar^2 (\mathbf{Q}-\mathbf{k})^2}{2m_e'}  - \frac{\hbar^2 k^2}{2m_x} + i\delta}, \nonumber \\
G'_0(\mathbf{k})  &=&  \frac{1}{E - \frac{\hbar^2 (\mathbf{Q}-\mathbf{k})^2}{2m_e}  - \frac{\hbar^2 k^2}{2m'_x} + \Delta_{xe} + i\delta}.
 \label{eq:G0d}
\end{eqnarray}
$\Delta_{xe}$ takes into account the difference in binding energies of excitons $\phi$ and $\phi'$ as well as a possible conduction-band energy difference. For example, if we wish to evaluate the intravalley trion in WSe$_2$ or WS$_2$ monolayers, then $G_0$ and $G'_0$ in Eq.~(\ref{eq:G0d}) correspond to the left and right diagrams of  Fig.~\ref{fig:M}(b), respectively. $G_0$ comprises a bright exciton and electron in the bottom valley whereas $G'_0$ comprises a dark exciton and electron in the top valley. Using experimental results that the dark exciton resides 40~meV below the bright one \cite{Zhang_NatNano17,Zhou_NatNano17,Wang_PRL17,He_NatComm20} and the conduction-band spin splitting energy is 12~meV \cite{Kapuscinski_NatComm21,Ren_PRB23}, we should use $\Delta_{xe}=28$~meV in such a case.

The second step in the hierarchy, shown in Fig.~\ref{fig:BSEd}(c)-(d),  is to incorporate the exchange interaction and calculate the final Green's function $g$ from the intermediate one $G$. The BSEs are formally written as
\begin{eqnarray}
g(\mathbf{k})  &=&  G(\mathbf{k}) + G(\mathbf{k}) \sum_{\mathbf{p}}    \lambda(\mathbf{Q},\mathbf{k},\mathbf{p})  g(\mathbf{p}), \nonumber \\
g'(\mathbf{k})  &=&  G'(\mathbf{k}) + G'(\mathbf{k}) \sum_{\mathbf{p}}  \lambda'(\mathbf{Q},\mathbf{k},\mathbf{p})  g'(\mathbf{p}),
 \label{eq:BSEexc}
\end{eqnarray}
where
\begin{eqnarray}
\lambda(\mathbf{Q},\mathbf{k},\mathbf{p})  &=&  \sum_{\mathbf{q}} X(\mathbf{Q},\mathbf{k},\mathbf{q})G'(\mathbf{q})X'(\mathbf{Q},\mathbf{q},\mathbf{p}), \nonumber \\
\lambda'(\mathbf{Q},\mathbf{k},\mathbf{p})  &=&   \sum_{\mathbf{q}} X'(\mathbf{Q},\mathbf{k},\mathbf{q})G(\mathbf{q})X(\mathbf{Q},\mathbf{q},\mathbf{p}) .
 \label{eq:lam}
\end{eqnarray}

Our goal is to solve the Equation set (\ref{eq:BSEdir}) and (\ref{eq:BSEexc}) with the help of Eqs.~(\ref{eq:D}), (\ref{eq:X}), (\ref{eq:G0d}) and (\ref{eq:lam}), after which we can recognize if a bound state emerges as a resonance in the negative energy of part of the spectral function,
 \begin{eqnarray}
A_\text{d}(E,\mathbf{Q})  &\propto& -\mathcal{I}m \left \{ \sum_\mathbf{k} g(\mathbf{k}) \right \}.
 \label{eq:Ad}
\end{eqnarray}
If a bound state emerges, it should also appear at the same energy when calculated through $g'$. This property makes use of the sum rule,  $\sum_\mathbf{k} g(\mathbf{k}) = \sum_\mathbf{k} g'(\mathbf{k})$ at the spectral position of the trion.

%%%%%%%%%%%%%%%%%%%%%%%%%%%%%%%%%%%%%%%%%%%%%%%%%%%%%%%%%%
%%%%%%%%%%%%%%%%%%%%%%%%%%%%%%%%%%%%%%%%%%%%%%%%%%%%%%%%%%
\subsection{Numerical Scheme}

The calculation of the direct and exchange interactions in Eqs.~(\ref{eq:D})-(\ref{eq:X}) is made with $1s$ exciton states. When using the ideal 2D Coulomb potential (hydrogen model), we get $\phi_\mathbf{k}= \phi'_\mathbf{k}=\sqrt{8\pi}a_x/[1 + (ka_x)^2]^{3/2}$, where the Bohr radius is $a_x = \hbar^2 \epsilon_b / e^2 \mu_x$ with $\mu_x  = m_em_h/(m_e+m_h)$. When using other potential forms (discussed below), the exciton states are calculated numerically. %There are various numerical methods to calculate these states, and we use the stochastic variational method in this work \cite{SVM}.

Below, we show solutions of the BSE for idle trions ($Q=0$). This assumption reduces the computational effort by rendering the Green's functions isotropic, where $G_0(\mathbf{k}) = G_0(k)$, and the same holds for $G_0'$, $G$, $G'$, $g$ and $g'$. As a result, angular integrations due to the sums in Eqs.~(\ref{eq:BSEid}), (\ref{eq:BSEdir}) and  (\ref{eq:BSEexc}) only involve the interaction terms. Appendix~\ref{app:num} provides numerical details on the conversion of these equations to a matrix inversion problem.
%For example,
% \begin{eqnarray}
%\bar{D}(p,k) \!=\!  \int_0^{2\pi} \! d\theta D(\mathbf{p},\mathbf{k})\,\, , 
%\end{eqnarray}
%where $\theta$ is the angle between $\mathbf{p}$ and $\mathbf{k}$. Similar integrations are used to define $\bar{D}'$, $\bar{\lambda}$ and $\bar{\lambda}'$. Once these quantities are calculated, Eqs.~(\ref{eq:BSEdir}) and (\ref{eq:BSEexc}) can be solved straightforwardly by matrix inversion technique. 

To show that component exchange in exciton-electron systems is a universal property, not limited to specific details of the Coulomb potential, we have performed calculations with three types of dielectric screening, $\epsilon(q)$, in the Coulomb potential, $V_{\mathbf{q}} = 2\pi e^2/A\epsilon(q) q$,
\begin{equation}
  \epsilon(q) =
    \begin{cases}
      \epsilon_r & \text{2D},\\
      \epsilon_r +r_0 q & \text{RK},\\
      \frac{1}{2}\left[\frac{N_t(q)}{D_t(q)}+\frac{N_b(q)}{D_b(q)}\right] & 3\chi.
    \end{cases}   \label{eq:dielectric}    
\end{equation}
The first case is the aforementioned ideal 2D potential (hydrogen model) where $\epsilon_r$ is the average dielectric constant of the environment around the 2D semiconductor. The second model is the celebrated Rytova-Keldysh (RK) potential \cite{Rytova_MSU67,Keldysh_JETP79,Cudazzo_PRB2011}, which also considers the polarizability of the 2D semiconductor through the parameter $r_0$. The third model is the $3\chi$ potential which accounts for the geometry of a TMD monolayer \cite{VanTuan_PRB18}
\begin{eqnarray}\label{Eq:DiFv2def}
D_j(q) &=& 1+q\ell_- -q\ell_- (1+p_j)\text{e}^{-\frac{qd}{2}} - (1-q \ell_- ) p_j \text{e}^{-qd}, \nonumber \\
N_j(q) &=& \left(1+q\ell_-\right)\left(1+q\ell_+\right) \nonumber \\ & + & \left[\left(1-p_j\right)-\left(1+p_j\right)q\ell_+\right]q\ell_-\text{e}^{-\frac{qd}{2}} \nonumber \\
    &+& (1-q\ell_-)(1-q\ell_+ )p_j\text{e}^{-qd}, 
\end{eqnarray}
where $j=\{b,t\}$, $d$ is the thickness of the monolayer, and $\ell_\pm$ are the polarizabilities of the central sheet made transition-meta atoms and outer sheets made of chalcogen atoms. $p_j \equiv (\epsilon_j-1)/(\epsilon_j+1)$ is expressed in terms of the dielectric constants of the layers on top and below the TMD, $\epsilon_\text{t}$ and $\epsilon_\text{b}$. 

The parameters we use in this work correspond to WSe$_2$ monolayer that is encapsulated in hexagonal boron nitride. The dielectric screening parameters are $\epsilon_r=\epsilon_b=\epsilon_t=3.8$, $d=0.6$~nm, $\ell_\pm = 5.9d$ and $r_0=4.5$~nm \cite{VanTuan_PRB18}. The effective masses are $m_h=0.36m_0$ and $m_e=0.29m_0$ where $m_0$ is the free electron mass \cite{Kormanyos_2DMater15}. The broadening parameter in $G_0$ and $G_0'$ is $\delta = 1$~meV.  As we discuss below, the only two parameters we change are $\Delta_{xe}$ and $m_e'$. 

%%%%%%%%%%%%%%%%%%%%%%%%%%%%%%%%%%%%%%%%%%%%%%%
%%%%%%%%%%%%%%%%%%%%%%%%%%%%%%%%%%%%%%%%%%%%%%%
%%%%%%%%%%%%%%%%%%%%%%%%%%%%%%%%%%%%%%%%%%%%%%%
%%%%%%%%%%%%%%%%%%%%%%%%%%%%%%%%%%%%%%%%%%%%%%%
\section{results and discussion}\label{sec:res}

\begin{figure}[t!]
\includegraphics[width=8.5cm]{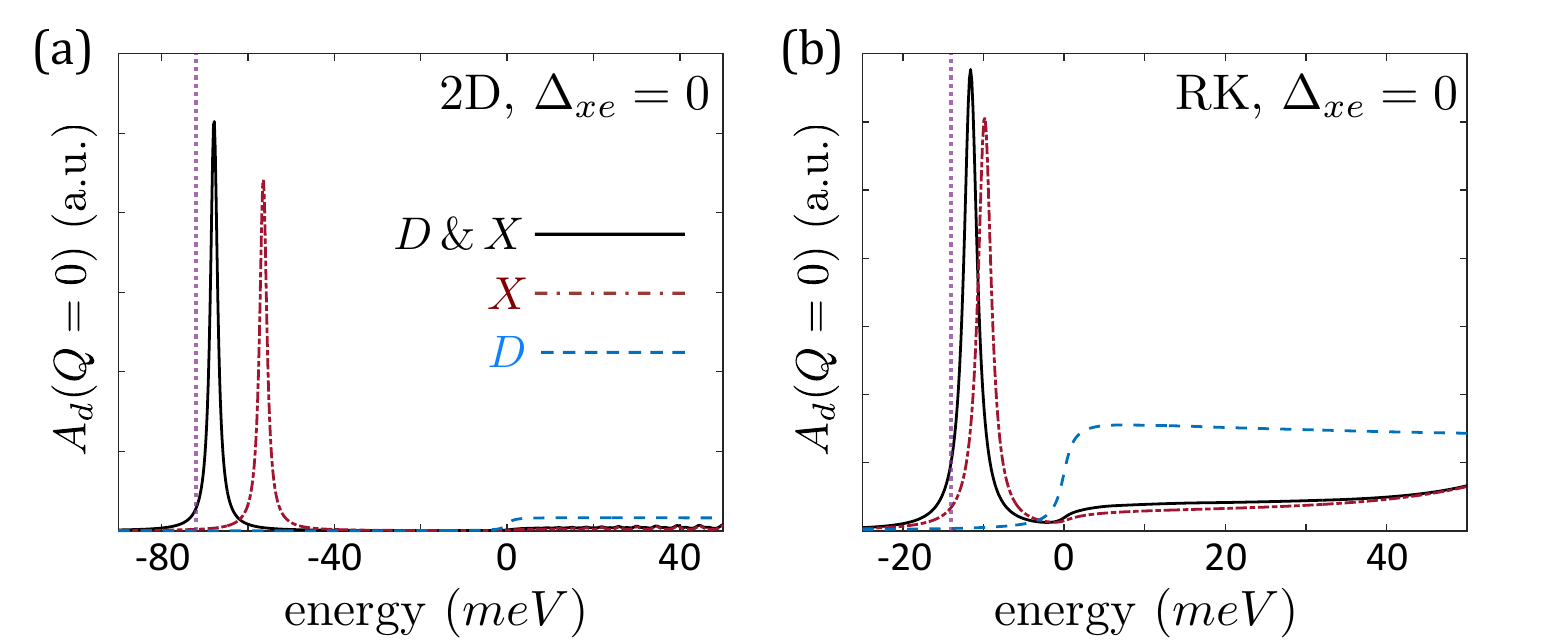}
\caption{Spectral functions of an exciton-electron system with  distinguishable electrons when $\Delta_{xe}=0$ and $m_e'=m_e$, calculated with the ideal 2D and RK potentials.} \label{fig:res0} 
\end{figure}

Figure~\ref{fig:res0} shows spectral functions of an exciton-electron system with two distinguishable electrons when $\Delta_{xe}=0$ and $m'_e=m_e$, calculated with the ideal 2D potential in Fig.~\ref{fig:res0}(a) and the RK potential in Fig.~\ref{fig:res0}(b). To understand the roles of the direct and exchange interactions, we have simulated three cases for each potential. The dashed lines are calculations that only involve the direct interaction, the dashed-dotted lines only involve the exchange interaction, and the solid lines involves both interactions. The vertical dotted lines at $-72$~meV in Fig.~\ref{fig:res0}(a) and at $-14$~meV in Fig.~\ref{fig:res0}(b) correspond to the binding energy of a trion when calculated as a three-body problem with the stochastic variational method \cite{VargaBook,Varga_PRC95}. The calculated binding energies with trion BSE are 3 to 4~meV smaller compared with those calculated through the three-body problem (difference between resonances of the solid lines and vertical dotted lines). We attribute this small deviation to the fact that the exciton binding energy is not infinite. That is, the trion's BSE model conforms to the solution of a three-body problem in the asymptotic limit that $a_x \rightarrow 0$. 

%%%%%%%%%%%%%%%%%%%%%%%%%%%%%%%%%%%%%%%%%%%%%%%%%%%%%%%%%%
%%%%%%%%%%%%%%%%%%%%%%%%%%%%%%%%%%%%%%%%%%%%%%%%%%%%%%%%%% 

\begin{figure}[t!]
\includegraphics[width=8.5cm]{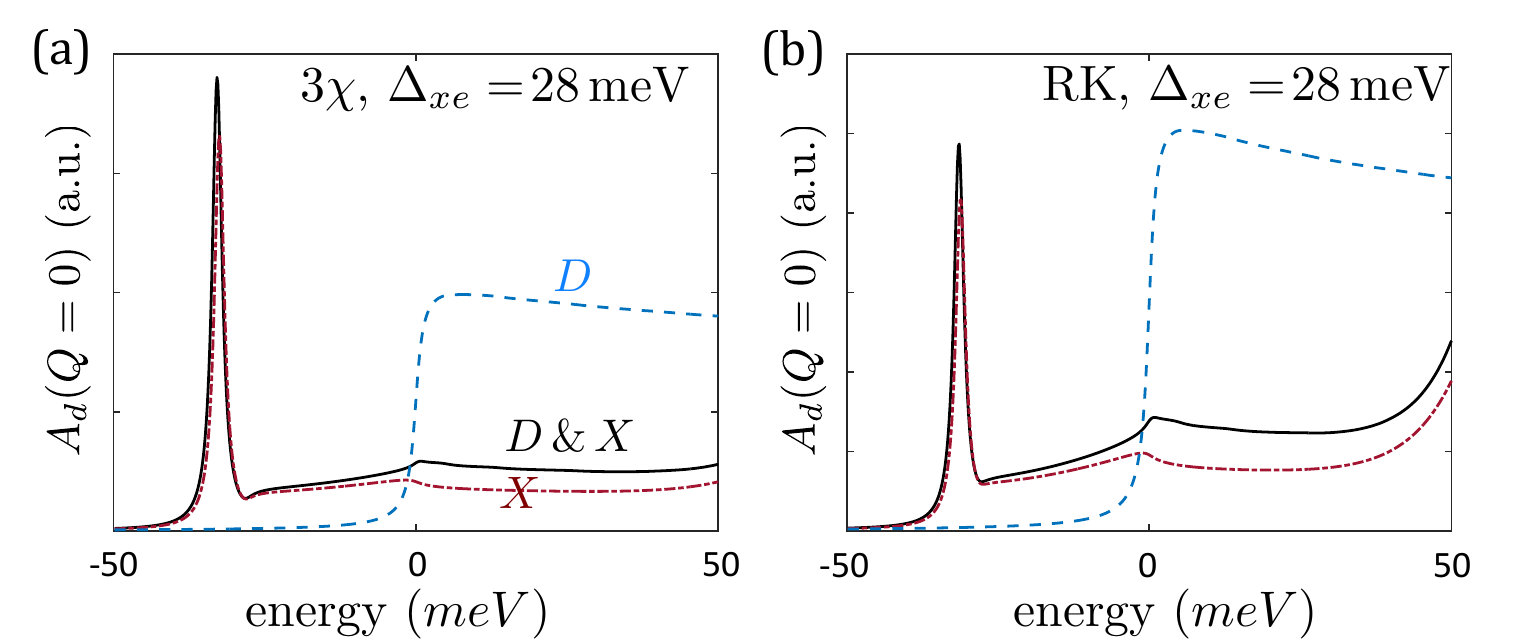}
\caption{Spectral functions of an exciton-electron system with  distinguishable electrons when $\Delta_{xe}=28$~meV and $m_e'=0.4m_0$, calculated with the $3\chi$ and RK potentials.} \label{fig:res28} 
\end{figure}

Figure~\ref{fig:res28} shows spectral functions of an exciton-electron system with two distinguishable electrons when $\Delta_{xe}=28$~meV and $m_e'=0.4m_0$, calculated with the $3\chi$ potential in Fig.~\ref{fig:res28}(a) and the RK potential in Fig.~\ref{fig:res28}(b). The motivation for using $\Delta_{xe}=28$~meV was mentioned after Eq.~(\ref{eq:G0d}), and the use of $m_e'=0.4m_0$ corresponds to the effective mass of an electron in the bottom valley of WSe$_2$ monolayer \cite{Kormanyos_2DMater15}. A noticeable qualitative difference between Figs.~\ref{fig:res0} and \ref{fig:res28} is that the direct interaction has no effect on the binding energy of the trion when $\Delta_{xe}$ is finite. Namely, the binding energy in Fig.~\ref{fig:res28} is the same with or without the direct interaction (the resonance positions of the dashed-dotted  and solid lines are identical). The reason for this behavior is that $G_0$ and $G_0'$ are tuned out of resonance when $\Delta_{xe}\neq 0$, such that the weak direct interaction cannot assist in the coupling of $\phi$ and $\phi'$. The weight of the direct interaction increases when the exciton-electron systems in $G_0$ and $G_0'$ resonate, as shown by the few meV difference between the trion peaks of the dashed-dotted  and solid lines in Fig.~\ref{fig:res0}.

The important finding of the analysis so far is that a bound trion state emerges because of component exchange. A bound trion state does not emerge when we solely consider the direct interaction (dashed-lines in Figs.~\ref{fig:res0} and \ref{fig:res28}). The spectral functions in this case show a step-function behavior around zero energy, which is simply the density-of-states of free particles in 2D. Namely, $A_\text{d}(E,Q=0) \propto \mu_T\theta(E)$, where $\mu_T  = m'_em_x/(m_x+m'_e)$ is the reduced mass of the unbound trion and $\theta(E)$ is the step function. When the exchange interaction is turned on, the spectral weight from small positive energies is transferred to the spectral region of the bound trion state. This behavior is recognized from the difference between the spectral functions at $E>0$ when the calculation is carried with and without the exchange interaction. Finally, we note that the ratio between the amplitudes of the trion resonance and step function region is a reflection of the spectral-weight transfer from the exciton-electron continuum ($E>0$) to the bound state (resonance). The stronger the Coulomb potential, the larger are the binding energy and spectral weight transfer.

Another implication of using a finite $\Delta_{xe}$ is that the trions binding energy is rendered less sensitive to the dielectric environment around the monolayer. Comparing Figs.~\ref{fig:res0}(b) and \ref{fig:res28}(b), the trion binding energy shifts from $-11.6$~meV when $\Delta_{xe}=0$  to $-31.5$~meV when $\Delta_{xe}=28$~meV. Because $\Delta_{xe}$ is solely governed by parameters of the monolayer \cite{Yang_PRB22}, this observation can explain why experiments show a relatively small change in the binding energy of trions when the dielectric environment around the monolayer changes. Namely, the spectral position of the trion's resonance is largely affected by $\Delta_{xe}$ and to a lesser effect by the dielectric environment.

%%%%%%%%%%%%%%%%%%%%%%%%%%%%%%%%%%%%%%%%%%%%%%%%%%%%%%%%%%
%%%%%%%%%%%%%%%%%%%%%%%%%%%%%%%%%%%%%%%%%%%%%%%%%%%%%%%%%%

\begin{figure}[t!] 
\centering
\includegraphics[width=6.5cm]{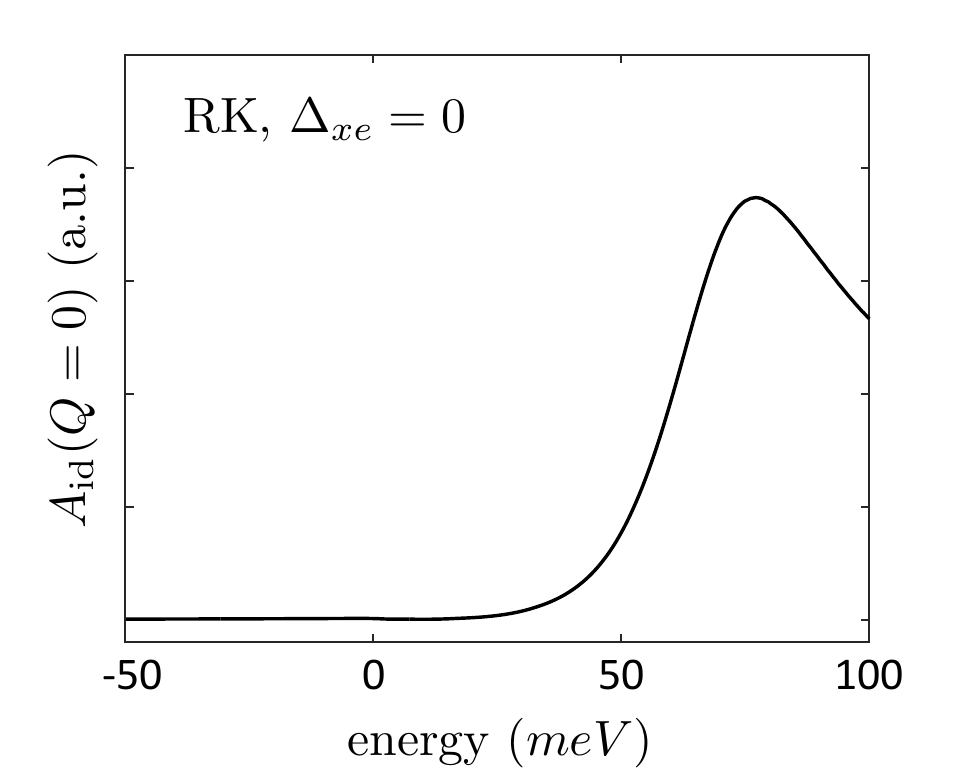}
\caption{ Spectral function of an exciton-electron system with indistinguishable electrons.} \label{fig:id} 
\end{figure}

Finally, we simulate an exciton-electron system with indistinguishable electrons through the solution of Eqs.~(\ref{eq:BSEid})-(\ref{eq:Aid}). Figure~\ref{fig:id} shows the resulting spectral function when using the RK potential. All the parameters are the same as in Fig.~\ref{fig:res0}(b) but the use of $\delta_d = 0$ in the exchange interaction. No bound state emerges at negative energies and this behavior persists regardless of the Coulomb potential we use (e.g., 2D or $3\chi$). Unlike the distinguishable case, the partition to direct and exchange contributions is meaningless because one cannot tell from the final state whether it was the exchange or direct interaction that took place. Figure~\ref{fig:id} shows that the spectral weight is transferred from small positive energies to larger positive energies. This behavior is attributed to the repulsive nature of the exciton-electron interaction when the electrons are indistinguishable. All in all, the spectral weight is transferred from small positive energies to form a bound state at negative energies when the component exchange is governed by attractive interaction ($\delta_d = 1$) and to larger positive energies when it is governed by repulsive interaction ($\delta_d = 0$). The stronger the Coulomb potential is, the transfer is carried to farther energies.
%%%%%%%%%%%%%%%%%%%%%%%%%%%%%%%%%%%%%%%%%%%%%%%
%%%%%%%%%%%%%%%%%%%%%%%%%%%%%%%%%%%%%%%%%%%%%%%
%%%%%%%%%%%%%%%%%%%%%%%%%%%%%%%%%%%%%%%%%%%%%%%
%%%%%%%%%%%%%%%%%%%%%%%%%%%%%%%%%%%%%%%%%%%%%%%
\section{Conclusions}\label{sec:sum}

We have shown that the short-range interaction between an exciton and electron is governed by the ability of the hole to switch between its electron partners (or an electron to switch between its hole partners in exciton-hole systems). This mechanism, coined component exchange, sheds light on the formation process of trions in photoluminescence and absorption measurements. The experimental conditions in these measurements are such that trions are formed when an exciton interacts with an electron rather than when a free hole interacts with two free electrons.  The importance of component exchange is further supported by its dominant contribution to the ultrafast energy relaxation of hot excitons when they interact with charge particles in lightly-doped semiconductors \cite{Yang_PRB22}. We have presented an effective set of Bethe-Salpeter Equations that one can use to calculate the trion binding energy without the use of fitting parameters.  Since the anti-commutation relations of the trion's constituent fermions are encoded in the exciton-electron interaction, a bound trion state naturally emerges when its particles are distinguishable. 
%%%%%%%%%%%%%%%%%%%%%%%%%%%%%%%%%%%%%%%%%%%%%%%%%%%%%%%%%%

\acknowledgments{ This work is supported by the Department of Energy, Basic Energy Sciences, Division of Materials Sciences and Engineering under Award No. DE-SC0014349.}
%%%%%%%%%%%%%%%%%%%%%%%%%%%%%%%%%%%%%%%%%%%%%%%%%%%%%%%%%%%%%%%%%%%%%%%%%%%%%%%
%%%%%%%%%%%%%%%%%%%%%%%%%%%%%%%%%%%%%%%%%%%%%%%%%%%%%%%%%%%%%%%%%%%%%%%%%%%%%%%
%%%%%%%%%%%%%%%%%%%%%%%%%%%%%%%%%%%%%%%%%%%%%%%%%%%%%%%%%%%%%%%%%%%%%%%%%%%%%%%
%%%%%%%%%%%%%%%%%%%%%%%%%%%%%%%%%%%%%%%%%%%%%%%%%%%%%%%%%%%%%%%%%%%%%%%%%%%%%%%

\appendix
%%%%%%%%%%%%%%%%%%%%%%%%%%%%%%%%%%%%%%%%%%%%%%%%%%%%%%%%%%%%%%%%%%%%%%%%%%%%%%%
%%%%%%%%%%%%%%%%%%%%%%%%%%%%%%%%%%%%%%%%%%%%%%%%%%%%%%%%%%%%%%%%%%%%%%%%%%%%%%%

\begin{widetext}
\section{Derivation of Eqs.~(\ref{eq:D}) and (\ref{eq:X})} \label{app:M}

\textbf{Indistinguishable electrons:} The state of an exciton with wavevector ${\bf k}$ and electron with wavevector ${\bf Q} - {\bf k}$ is written in second quantization as \cite{Ramon_PRB03} 
\begin{eqnarray}
| {\bf k},{\bf Q} \!-\! {\bf k} \rangle  \!=\!    \frac{1}{\sqrt{A}}    \sum_{{\bf k}_1} \phi^*_{\eta {\bf k} + {\bf k}_1  }   \psi^*_{{\bf Q} - {\bf k}} c^\dagger_{-{\bf k}_1} \!d^\dagger_{{\bf k} + {\bf k}_1} \!
  c^\dagger_{{\bf Q} - {\bf k}}  |0\rangle. \qquad \label{eq:state}
\end{eqnarray} 
$\psi_{\bf k}$ ($\phi_{\bf k}$) is the envelope function of the electron (exciton) in $\mathbf{k}$-space, and $|0\rangle$ is the ground state with no holes in the valence band and no electrons in the conduction band. In the absence of magnetic field, $\psi_{\bf k}=1$ due to translation symmetry. Therefore, we omit $\psi$ terms in what follows.  $c^\dagger_\mathbf{k}$ ($d^\dagger_\mathbf{k}$) are creation operators of an electron (hole). The interaction of an electron with the electron and hole of the exciton reads
\begin{eqnarray}
V_{ee} &=& \frac{1}{2} \sum_{\bf q,k',k''}  V_{\bf q}  c^\dagger_{{\bf k'+q}}  c^\dagger_{{\bf k''-q}}   c_{\bf k''}  c_{\bf k'} \,,\nonumber \\
V_{eh} &=& - \sum_{\bf q,k',k''}  V_{\bf q}  c^\dagger_{{\bf k'+q}}  d^\dagger_{{\bf k''-q}}   d_{\bf k''}  c_{\bf k'},
\end{eqnarray}
where the 1/2 factor is needed to avoid double counting due to indistinguishability of the electrons. $c_\mathbf{k}$ ($d_\mathbf{k}$) are annihilation operators of an electron (hole). Using Fermion anticommutation,  $   b^\dagger_{{\bf k}_1} b_{{\bf k}_2} +  b_{{\bf k}_2} b^\dagger_{{\bf k}_1}  =  \delta_{{{\bf k}_1}, {\bf k}_2}$,  and the relation $b_{\bf k}  | 0 \rangle = 0 $ where $b=\{c,d\}$, we can derive that
\begin{eqnarray}
V_{ee} | {\bf k},{\bf Q} - {\bf k} \rangle  &=& \frac{1}{2\sqrt{A}} \sum_{ {\bf q}, {\bf k}', {\bf k}'',{\bf k}_1 }  V_{\bf q} \phi^*_{\eta {\bf k} + {\bf k}_1  } c^\dagger_{{\bf k}'+{\bf q}}  c^\dagger_{{\bf k}''-{\bf q}}   c_{{\bf k}''}  c_{{\bf k}'}   c^\dagger_{-{\bf k}_1} d^\dagger_{{\bf k} + {\bf k}_1}  c^\dagger_{{\bf Q} - {\bf k}}   \left| 0 \right\rangle \nonumber \\
&=&  \frac{1}{2\sqrt{A}} \sum_{ {\bf q}, {\bf k}', {\bf k}'',{\bf k}_1 }    V_{\bf q} \phi^*_{\eta {\bf k} + {\bf k}_1  }    c^\dagger_{{\bf k}'+{\bf q}}  d^\dagger_{{\bf k} + {\bf k}_1}   c^\dagger_{{\bf k}''-{\bf q}}   \left( \delta_{{\bf k}'', {\bf Q} - {\bf k}}    \delta_{{\bf k}',-{\bf k}_1} -  \delta_{{\bf k}'', -{\bf k}_1}     \delta_{{\bf k}',{\bf Q} - {\bf k} } \right) \left| 0 \right\rangle \nonumber \\
&=&  \frac{1}{2\sqrt{A}} \sum_{ {\bf q}, {\bf k}_1 }   V_{\bf q} \phi^*_{\eta {\bf k} + {\bf k}_1  }   \left( c^\dagger_{{\bf q}-{\bf k}_1}  d^\dagger_{{\bf k} + {\bf k}_1}   c^\dagger_{{\bf Q}-{\bf k}-{\bf q}}  - c^\dagger_{{\bf Q}-{\bf k}+{\bf q}}  d^\dagger_{{\bf k} + {\bf k}_1}   c^\dagger_{-{\bf k}_1-{\bf q}}    \right)   \left| 0 \right\rangle \nonumber \\
 &=& \frac{1}{\sqrt{A}} \sum_{ {\bf q}, {\bf k}_1 }    V_{\bf q} \phi^*_{\eta {\bf k} + {\bf k}_1  }    c^\dagger_{{\bf q}-{\bf k}_1}  d^\dagger_{{\bf k} + {\bf k}_1}   c^\dagger_{{\bf Q}-{\bf k}-{\bf q}}\left| 0 \right\rangle.
 \label{eq:Vee1}
\end{eqnarray} 
 The last transition is possible because $V_{\bf q}=V_{-{\bf q}}$.  Using Eqs.~(\ref{eq:state}) and (\ref{eq:Vee1}), the electron-electron interaction part reads
 \begin{eqnarray}
\langle {\bf p},{\bf Q}' - {\bf p} |V_{ee} | {\bf k},{\bf Q} - {\bf k} \rangle  &=& \frac{1}{A}  \sum_{ {\bf q}, {\bf k}_1,{\bf p}_1 }    V_{\bf q} \phi_{\eta {\bf p} + {\bf p}_1  }  \phi^*_{\eta {\bf k} + {\bf k}_1  }  
\langle 0 | c_{{\bf Q}' - {\bf p}} d_{{\bf p} + {\bf p}_1} c_{-{\bf p}_1}   c^\dagger_{{\bf q}-{\bf k}_1}  d^\dagger_{{\bf k} + {\bf k}_1}   c^\dagger_{{\bf Q}-{\bf k}-{\bf q}} | 0 \rangle.  \nonumber \\
&=& \frac{1}{A}  \sum_{ {\bf q}, {\bf k}_1,{\bf p}_1 }    V_{\bf q} \phi_{\eta {\bf p} + {\bf p}_1  }  \phi^*_{\eta {\bf k} + {\bf k}_1  }  \delta_{{\bf p}_1 ,{\bf k} - {\bf p} + {\bf k}_1 } \left( \delta_{{\bf p}_1, {\bf k}_1 - {\bf q}}  \delta_{{\bf Q}' - {\bf p} , {\bf Q}-{\bf k}-{\bf q}} - \delta_{{\bf p}_1,{\bf k} - {\bf Q} +{\bf q}}  \delta_{{\bf Q}' - {\bf p} , {\bf q}-{\bf k}_1} \right) \qquad \nonumber \\
&=& \frac{\delta_{{\bf Q}' , {\bf Q}}}{A}  \sum_{ {\bf q}, {\bf k}_1, {\bf p}_1 }    V_{\bf q} \phi_{\eta {\bf p} +  {\bf p}_1  }  \phi^*_{\eta {\bf k} + {\bf k}_1  }  \left( \delta_{{\bf p}_1 ,{\bf k} - {\bf p} + {\bf k}_1 } \delta_{{\bf q}, {\bf p} - {\bf k}}  - \delta_{{\bf p}_1 , {\bf k} - {\bf Q} +{\bf q}} \delta_{{\bf k}_1,{\bf p} - {\bf Q} +{\bf q} }   \right) \nonumber \\
&=& \frac{\delta_{{\bf Q}' , {\bf Q}}}{A}  \sum_{ {\bf k}_1 }    V_{{\bf p} - {\bf k}} \phi_{\bar{\eta} {\bf p} + {\bf k} + {\bf k}_1  }  \phi^*_{\eta {\bf k} + {\bf k}_1  }  - \frac{\delta_{{\bf Q}' , {\bf Q}}}{A}  \sum_{ {\bf q} }   V_{\bf q} \phi_{\eta {\bf p} + {\bf k}  - {\bf Q}  +{\bf q} }  \phi^*_{\eta {\bf k} +{\bf p} - {\bf Q}  +{\bf q} }   \nonumber \\
&=& \frac{\delta_{{\bf Q}' , {\bf Q}}}{A}  \sum_{ {\bf q} }    V_{{\bf k} - {\bf p}}\, \phi_{\bar{\eta} {\bf p} + {\bf q}   } \, \phi^*_{\bar{\eta} {\bf k} + {\bf q}  }  - \frac{\delta_{{\bf Q}' , {\bf Q}}}{A}  \sum_{ {\bf q} }   V_{\bf q} \,\phi_{\eta {\bf p} + {\bf k} -{\bf Q} - {\bf q}  } \,  \phi^*_{\eta {\bf k} +{\bf p}-{\bf Q} - {\bf q}  }  \,\,\,  ,
 \label{eq:Vee2}
\end{eqnarray} 
where $\bar{\eta}=\eta-1$. To get the last line, the integration variable of the first sum is changed ${\bf k}_1 \rightarrow {\bf q} - {\bf k}$, followed by ${\bf q} \rightarrow -{\bf q}$ in both sums and using the fact that $V_{\bf q}=V_{-{\bf q}}$. Repeating the analysis for the electron-hole part, we get
\begin{eqnarray}
V_{eh} | {\bf k},{\bf Q} - {\bf k} \rangle  &=& - \frac{1}{\sqrt{A}} \sum_{ {\bf q}, {\bf k}_1 }    V_{\bf q} \phi^*_{\eta {\bf k} + {\bf k}_1  }    c^\dagger_{-{\bf k}_1}  d^\dagger_{{\bf k} + {\bf k}_1 + {\bf q}}   c^\dagger_{{\bf Q}-{\bf k}-{\bf q}}\left| 0 \right\rangle,
 \label{eq:Veh1}
\end{eqnarray} 
and
\begin{eqnarray}
\langle {\bf p},{\bf Q}' - {\bf p} |V_{eh} | {\bf k},{\bf Q} - {\bf k} \rangle  &=& - \frac{1}{A}  \sum_{ {\bf q}, {\bf k}_1,{\bf p}_1 }    V_{\bf q} \phi_{\eta {\bf p} + {\bf p}_1  }  \phi^*_{\eta {\bf k} + {\bf k}_1  }  
\langle 0 | c_{{\bf Q}' - {\bf p}} d_{{\bf p} + {\bf p}_1} c_{-{\bf p}_1}   c^\dagger_{-{\bf k}_1}  d^\dagger_{{\bf k} + {\bf k}_1 + {\bf q}}   c^\dagger_{{\bf Q}-{\bf k}-{\bf q}} | 0 \rangle.  \nonumber \\
&=&- \frac{1}{A}  \sum_{ {\bf q}, {\bf k}_1,{\bf p}_1 }    V_{\bf q} \phi_{\eta {\bf p} + {\bf p}_1  }  \phi^*_{\eta {\bf k} + {\bf k}_1  }  \delta_{{\bf p}_1 - {\bf k}_1 ,{\bf k}  - {\bf p} + {\bf q}} \left( \delta_{{\bf p}_1, {\bf k}_1 }  \delta_{{\bf Q}' - {\bf p} , {\bf Q}-{\bf k}-{\bf q}} - \delta_{{\bf p}_1,{\bf k} - {\bf Q} +{\bf q}}  \delta_{{\bf Q}' - {\bf p} , -{\bf k}_1} \right) \qquad \nonumber \\
&=& \frac{\delta_{{\bf Q}' , {\bf Q}}}{A}  \sum_{ {\bf q}, {\bf k}_1, {\bf p}_1}    V_{\bf q} \phi_{\eta {\bf p} +  {\bf p}_1  }  \phi^*_{\eta {\bf k} + {\bf k}_1  }  \left(  \delta_{{\bf p}_1,{\bf k} - {\bf Q} +{\bf q}}  \delta_{{\bf k}_1,{\bf p} - {\bf Q}}  -  \delta_{{\bf p}_1, {\bf k}_1 } \delta_{{\bf q}, {\bf p} - {\bf k}} \right) \nonumber \\
&=&  \frac{\delta_{{\bf Q}' , {\bf Q}}}{A}  \sum_{ {\bf q} }   V_{\bf q} \phi_{\eta {\bf p} + {\bf k}  - {\bf Q}  +{\bf q} }  \phi^*_{\eta {\bf k} +{\bf p} - {\bf Q}  }  - \frac{\delta_{{\bf Q}' , {\bf Q}}}{A}  \sum_{ {\bf k}_1 }    V_{{\bf p} - {\bf k}} \phi_{\eta {\bf p} + {\bf k}_1  }  \phi^*_{\eta {\bf k} + {\bf k}_1  }   \nonumber \\
&=&  \frac{\delta_{{\bf Q}' , {\bf Q}}}{A}  \sum_{ {\bf q} }   V_{\bf q} \,\phi_{\eta {\bf p} + {\bf k} -{\bf Q} - {\bf q}  } \,  \phi^*_{\eta {\bf k} +{\bf p}- {\bf Q}  }    -  \frac{\delta_{{\bf Q}' , {\bf Q}}}{A}  \sum_{ {\bf q} }    V_{{\bf k} - {\bf p}}\, \phi_{\eta {\bf p} + {\bf q}   } \, \phi^*_{\eta {\bf k} + {\bf q}  } .
 \label{eq:Veh2}
 \end{eqnarray} 
Combining Eqs.~(\ref{eq:Vee2}) and (\ref{eq:Veh2}), we arrive at Eqs.~(\ref{eq:M})-(\ref{eq:X})  for the case that the electrons are indistinguishable ($\delta_d=0$, $\eta = \eta'$ and $\phi'_{\bf k}=\phi_{\bf k}$),
\begin{eqnarray}
 M({\bf Q}, {\bf k},  {\bf p}) &=&  \langle {\bf p},{\bf Q} - {\bf p} |V_{\text{ee}}+V_{\text{eh}} | {\bf k},{\bf Q} - {\bf k} \rangle   \label{eq:Vid}\\ 
 &=& \frac{V_{{\bf k} - {\bf p}}}{A}  \sum_{\bf q}   \left[    \phi_{\bar{\eta} {\bf p} + {\bf q}   } \, \phi^*_{\bar{\eta} {\bf k} + {\bf q}  }   -    \phi_{\eta {\bf p} + {\bf q}   }  \, \phi^*_{\eta {\bf k} + {\bf q}  }\right] \,\, + \frac{1}{A}  \sum_{\bf q} V_{\bf q} \phi_{\eta {\bf p} + {\bf k} -{\bf Q} - {\bf q}  } \left[ \phi^*_{\eta {\bf k} +{\bf p}- {\bf Q}  }  -  \phi^*_{\eta {\bf k} +{\bf p}-{\bf Q} - {\bf q}  } \right] \nonumber .
 \end{eqnarray} 
$\\$
\textbf{Distinguishable electrons:} We repeat the analysis and first rewrite the exciton-electron state as
\begin{eqnarray}
| {\bf k},{\bf Q} \!-\! {\bf k} \rangle  \!=\!    \frac{1}{\sqrt{A}}    \sum_{{\bf k}_1} \phi^*_{\eta {\bf k} + {\bf k}_1  }   \psi^*_{{\bf Q} - {\bf k}} c^\dagger_{-{\bf k}_1} \!d^\dagger_{{\bf k} + {\bf k}_1} \!
  e^\dagger_{{\bf Q} - {\bf k}}  |0\rangle, \qquad \label{eq:state_d}
\end{eqnarray} 
where $e_\mathbf{k}\dagger$ is creation operator of an electron that is distinguishable from the one created by $c_\mathbf{k}\dagger$ (e.g., with opposite spins and/or different valleys). As before, $\psi_{\bf k}=1$ due to translation symmetry and it is omitted hereafter. Next, the interaction of the distinguishable electron with the electron and hole of the exciton are rewritten as
\begin{eqnarray}
V_{ee} + V_{eh} &=&  \sum_{\bf q,k',k''}  V_{\bf q}  \left( c^\dagger_{{\bf k'+q}}  e^\dagger_{{\bf k''-q}}   e_{\bf k''}  c_{\bf k'} -  e^\dagger_{{\bf k'+q}}  d^\dagger_{{\bf k''-q}}   d_{\bf k''}  e_{\bf k'} \right).
\end{eqnarray}
$e_\mathbf{k}$ is the annihilation operator of the distinguishable electron. Using Fermion anticommutation,  $   b^\dagger_{{\bf k}_1} b_{{\bf k}_2} +  b_{{\bf k}_2} b^\dagger_{{\bf k}_1}  =  \delta_{{{\bf k}_1}, {\bf k}_2}$,  and the relation $b_{\bf k}  | 0 \rangle = 0 $ where $b=\{c,d,e\}$, we derive that
\begin{eqnarray}
&V_{ee}& | {\bf k},{\bf Q} - {\bf k} \rangle  = \frac{1}{\sqrt{A}} \sum_{ {\bf q}, {\bf k}', {\bf k}'',{\bf k}_1 }  V_{\bf q} \phi^*_{\eta {\bf k} + {\bf k}_1  } c^\dagger_{{\bf k}'+{\bf q}}  e^\dagger_{{\bf k}''-{\bf q}}   e_{{\bf k}''}  c_{{\bf k}'}   c^\dagger_{-{\bf k}_1} d^\dagger_{{\bf k} + {\bf k}_1}  e^\dagger_{{\bf Q} - {\bf k}}   \left| 0 \right\rangle   \label{eq:Vee1d} \\
&=&  \frac{1}{\sqrt{A}} \sum_{ {\bf q}, {\bf k}', {\bf k}'',{\bf k}_1 }    V_{\bf q} \phi^*_{\eta {\bf k} + {\bf k}_1  }    c^\dagger_{{\bf k}'+{\bf q}}  d^\dagger_{{\bf k} + {\bf k}_1}   e^\dagger_{{\bf k}''-{\bf q}}  \delta_{{\bf k}'', {\bf Q} - {\bf k}}    \delta_{{\bf k}',-{\bf k}_1}   \left| 0 \right\rangle =  \frac{1}{\sqrt{A}} \sum_{ {\bf q}, {\bf k}_1 }   V_{\bf q} \phi^*_{\eta {\bf k} + {\bf k}_1  }   c^\dagger_{{\bf q}-{\bf k}_1}  d^\dagger_{{\bf k} + {\bf k}_1}   e^\dagger_{{\bf Q}-{\bf k}-{\bf q}}     \left| 0 \right\rangle ,\nonumber
\end{eqnarray} 
and
\begin{eqnarray}
&V_{eh}& | {\bf k},{\bf Q} - {\bf k} \rangle = - \frac{1}{\sqrt{A}} \sum_{ {\bf q}, {\bf k}', {\bf k}'',{\bf k}_1 } \! V_{\bf q} \phi^*_{\eta {\bf k} + {\bf k}_1  }   e^\dagger_{{\bf k}'+{\bf q}}  d^\dagger_{{\bf k}''-{\bf q}}   d_{{\bf k}''}   e_{{\bf k}'}  c^\dagger_{-{\bf k}_1} d^\dagger_{{\bf k} + {\bf k}_1}  e^\dagger_{{\bf Q} - {\bf k}}   \left| 0 \right\rangle  \label{eq:Veh1d} \\
&=&  -\frac{1}{\sqrt{A}} \sum_{ {\bf q}, {\bf k}', {\bf k}'',{\bf k}_1 } \!  V_{\bf q} \phi^*_{\eta {\bf k} + {\bf k}_1  }        e^\dagger_{{\bf k}'+{\bf q}}  d^\dagger_{{\bf k}''-{\bf q}}  c^\dagger_{- {\bf k}_1}  \delta_{{\bf k}',  {\bf Q}- {\bf k} }\delta_{{\bf k}'',{\bf k} + {\bf k}_1}       \left| 0 \right\rangle =  -\frac{1}{\sqrt{A}} \sum_{ {\bf q}, {\bf k}_1 }  V_{\bf q} \, \phi^*_{\eta {\bf k} + {\bf k}_1  }      e^\dagger_{{\bf Q}- {\bf k}+{\bf q}}  d^\dagger_{ {\bf k} + {\bf k}_1 - {\bf q}}  c^\dagger_{- {\bf k}_1}          \left| 0 \right\rangle . \nonumber
\end{eqnarray} 
To write the interaction matrix elements when the electrons are distinguishable,  we need to know whether the final state includes an exciton with the same or different  electron as in the initial state,
\begin{eqnarray}
_d\langle {\bf p},{\bf Q}' \!-\! {\bf p} |  &=&   \!  \frac{1}{\sqrt{A}}    \sum_{{\bf p}_1} \phi_{\eta {\bf p} + {\bf p}_1  }  \langle 0 |  e_{{\bf Q}' - {\bf p}}  d_{{\bf p} + {\bf p}_1}  c_{-{\bf p}_1} 
\,\,    ,  \nonumber \\
  _x\langle {\bf p},{\bf Q}' \!-\! {\bf p} |  &=&  \!  \frac{1}{\sqrt{A}}    \sum_{{\bf p}_1} \phi'_{\eta' {\bf p} + {\bf p}_1  }  \langle 0 |    e_{-{\bf p}_1}  d_{{\bf p} + {\bf p}_1}  
  c_{{\bf Q}' - {\bf p}}  
. \qquad \,\,\,\,\,\,  \label{eq:state_d}
\end{eqnarray} 
Connecting the ket states in Eqs.~(\ref{eq:Vee1d})-(\ref{eq:Veh1d}) with the first bra state in Eq.~(\ref{eq:state_d}), the direct interaction is given by 
\begin{eqnarray}
&\langle&\!\! {\bf p},{\bf Q}' \!-\! {\bf p} | V_{ee} + V_{eh}| {\bf k},{\bf Q} - {\bf k}\, \rangle_d =  \frac{1}{A}    \sum_{{\bf q}, {\bf k}_1,{\bf p}_1} V_{\bf q}  \phi_{\eta {\bf p} + {\bf p}_1  } \phi^*_{\eta {\bf k} + {\bf k}_1  }  \left( \langle 0 |  e_{{\bf Q}' - {\bf p}}  d_{{\bf p} + {\bf p}_1}  c_{-{\bf p}_1}       c^\dagger_{{\bf q}-{\bf k}_1}  d^\dagger_{{\bf k} + {\bf k}_1}   e^\dagger_{{\bf Q}-{\bf k}-{\bf q}}     \left| 0 \right\rangle  \right. \nonumber \\  && \qquad  \qquad  \qquad  \qquad\qquad  \qquad \qquad  \qquad \qquad \qquad  \qquad \qquad  \qquad \left.  - \langle 0 |  e_{{\bf Q}' - {\bf p}}  d_{{\bf p} + {\bf p}_1}  c_{-{\bf p}_1}       e^\dagger_{{\bf Q}- {\bf k}+{\bf q}}  d^\dagger_{ {\bf k} + {\bf k}_1 - {\bf q}}  c^\dagger_{- {\bf k}_1}     \left| 0 \right\rangle \right) \nonumber \\
&=&  \frac{1}{A}    \sum_{{\bf q}, {\bf k}_1,{\bf p}_1} V_{\bf q}  \phi_{\eta {\bf p} + {\bf p}_1  } \phi^*_{\eta {\bf k} + {\bf k}_1  } \left(  \delta_{{\bf p}_1,{\bf k} - {\bf p} + {\bf k}_1}       \delta_{{\bf p}_1,{\bf k}_1 - {\bf q} }        \delta_{{\bf Q}' , {\bf Q}-{\bf k}+{\bf p}-{\bf q}}  -  \delta_{{\bf p}_1,{\bf k} - {\bf p} + {\bf k}_1  - {\bf q}} \delta_{{\bf p}_1,{\bf k}_1  }        \delta_{{\bf Q}' , {\bf Q}-{\bf k}+{\bf p}+{\bf q}} \right) \nonumber \\ 
\nonumber \\ & =&  \frac{\delta_{{\bf Q},{\bf Q}'}}{A}  \left[  \sum_{{\bf k}_1} V_{{\bf p} - {\bf k}}  \phi_{\bar{\eta} {\bf p} + {\bf k} + {\bf k}_1  } \phi^*_{\eta {\bf k} + {\bf k}_1  }  -  \sum_{{\bf k}_1} V_{{\bf k} - {\bf p}}  \phi_{\eta {\bf p} + {\bf k}_1  } \phi^*_{\eta {\bf k} + {\bf k}_1  } \right]  \nonumber \\ 
&=&  \delta_{{\bf Q},{\bf Q}'} \frac{V_{{\bf k} - {\bf p}}}{A}    \sum_{{\bf q}}   \left [ \phi_{\bar{\eta} {\bf p} + {\bf q}   } \phi^*_{\bar{\eta} {\bf k} + {\bf q}  }  -   \phi_{\eta {\bf p} + {\bf q}   } \phi^*_{\eta {\bf k} + {\bf q}  }      \right] \,. \label{eq:Dd}
 %%%%%%%%
\end{eqnarray} 
To get the first sum in the last line, the integration variable is changed ${\bf k}_1 \rightarrow {\bf q} - {\bf k}$ and we assign $\bar{\eta}=\eta-1$. 

Connecting the ket states in Eqs.~(\ref{eq:Vee1d})-(\ref{eq:Veh1d}) with the second bra state in Eq.~(\ref{eq:state_d}), the component  exchange interaction is given by 
\begin{eqnarray}
\,\nonumber  \\ \nonumber  \\
&\langle&\! \!{\bf p},{\bf Q}' \!-\! {\bf p} | V_{ee} +  V_{eh} | {\bf k},{\bf Q} - {\bf k}\, \rangle_x = \frac{1}{A}    \sum_{{\bf q}, {\bf k}_1,{\bf p}_1} V_{\bf q}  \phi'_{\eta' {\bf p} + {\bf p}_1  }  \phi^*_{\eta {\bf k} + {\bf k}_1  }  \left(  \langle 0 |    e_{-{\bf p}_1}  d_{{\bf p} + {\bf p}_1} c_{{\bf Q}' - {\bf p}}   c^\dagger_{{\bf q}-{\bf k}_1}  d^\dagger_{{\bf k} + {\bf k}_1}   e^\dagger_{{\bf Q}-{\bf k}-{\bf q}}     \left| 0 \right\rangle \right. \nonumber \\  &&  \qquad  \qquad  \qquad  \qquad\qquad  \qquad \qquad  \qquad \qquad  \qquad \qquad  \qquad \qquad   \left.  - \langle 0 |  e_{-{\bf p}_1}  d_{{\bf p} + {\bf p}_1} c_{{\bf Q}' - {\bf p}}        e^\dagger_{{\bf Q}- {\bf k}+{\bf q}}  d^\dagger_{ {\bf k} + {\bf k}_1 - {\bf q}}  c^\dagger_{- {\bf k}_1}     \left| 0 \right\rangle \right) \nonumber \\
&=&  \frac{1}{A}    \sum_{{\bf q}, {\bf k}_1,{\bf p}_1} V_{\bf q}  \phi'_{\eta' {\bf p} + {\bf p}_1  }  \phi^*_{\eta {\bf k} + {\bf k}_1  } \left(       \delta_{{\bf p}_1,{\bf k}  -  {\bf Q} + {\bf q}}      \delta_{{\bf p}_1,{\bf k} - {\bf p} + {\bf k}_1}    \delta_{{\bf Q}' , {\bf p}+{\bf q}-{\bf k}_1}   -   \delta_{{\bf p}_1,{\bf k}-{\bf Q} -{\bf q} }    \delta_{{\bf p}_1,{\bf k} - {\bf p} + {\bf k}_1  - {\bf q}}    \delta_{{\bf Q}' , {\bf p}-{\bf k}_1} \right) \nonumber \\ 
\nonumber \\ & =&  \frac{\delta_{{\bf Q},{\bf Q}'}}{A}  \left[  \sum_{{\bf q}} V_{{\bf q} }  \phi'_{\eta' {\bf p} + {\bf k} - {\bf Q} + {\bf q}   }  \phi^*_{\eta {\bf k} + {\bf p}  -  {\bf Q} + {\bf q} }   -  \sum_{{\bf q}} V_{{\bf q}}  \phi'_{\eta' {\bf p} + {\bf k} - {\bf Q} - {\bf q}  }  \phi^*_{\eta {\bf k} + {\bf p} - {\bf Q} } \right]  \nonumber \\ 
&=& - \frac{\delta_{{\bf Q},{\bf Q}'}}{A}    \sum_{{\bf q}} V_{{\bf q} }  \phi'_{\eta' {\bf p} + {\bf k} - {\bf Q} - {\bf q}  } \left [   \phi^*_{\eta {\bf k} + {\bf p} - {\bf Q} } -  \phi^*_{\eta {\bf k} + {\bf p}  -  {\bf Q} - {\bf q} }   \right]  \,.  \label{eq:Ed}
 %%%%%%%%
\end{eqnarray} 
Combining Eqs.~(\ref{eq:Dd}) and (\ref{eq:Ed}), we arrive at Eqs.~(\ref{eq:M})-(\ref{eq:X})  for the case that the electrons are distinguishable ($\delta_d=1$),
\begin{eqnarray}
 M({\bf Q}, {\bf k},  {\bf p})  &=& \frac{V_{{\bf k} - {\bf p}}}{A}  \sum_{\bf q}   \left[    \phi_{\bar{\eta} {\bf p} + {\bf q}   } \, \phi^*_{\bar{\eta} {\bf k} + {\bf q}  }   -    \phi_{\eta {\bf p} + {\bf q}   }  \, \phi^*_{\eta {\bf k} + {\bf q}  }\right] \,\, - \frac{1}{A}  \sum_{\bf q} V_{\bf q} \phi'_{\eta' {\bf p} + {\bf k} -{\bf Q} - {\bf q}  } \left[ \phi^*_{\eta {\bf k} +{\bf p}- {\bf Q}  }  -  \phi^*_{\eta {\bf k} +{\bf p}-{\bf Q} - {\bf q}  } \right] .\qquad
 \end{eqnarray} 
Comparing this result with Eq.~(\ref{eq:Vid}),  the direct interaction (first sum) is the same in the indistinguishable and distinguishable cases, whereas the exchange interaction (second sum) has opposite sign. Hence the use of $(-1)^{\delta_d}$ in Eq.~(\ref{eq:X}).
\end{widetext}

%%%%%%%%%%%%%%%%%%%%%%%%%%%%%%%%%%%%%%%%%%%%%%%%%%%%%%%%%%%%%%%%%%%%%%%%%%%%%%%
%%%%%%%%%%%%%%%%%%%%%%%%%%%%%%%%%%%%%%%%%%%%%%%%%%%%%%%%%%%%%%%%%%%%%%%%%%%%%%%
%%%%%%%%%%%%%%%%%%%%%%%%%%%%%%%%%%%%%%%%%%%%%%%%%%%%%%%%%%%%%%%%%%%%%%%%%%%%%%%
%%%%%%%%%%%%%%%%%%%%%%%%%%%%%%%%%%%%%%%%%%%%%%%%%%%%%%%%%%%%%%%%%%%%%%%%%%%%%%%
\section{BSE derivation} \label{app:BSE}

\begin{figure}[t] 
\centering
\includegraphics[width=6.5cm]{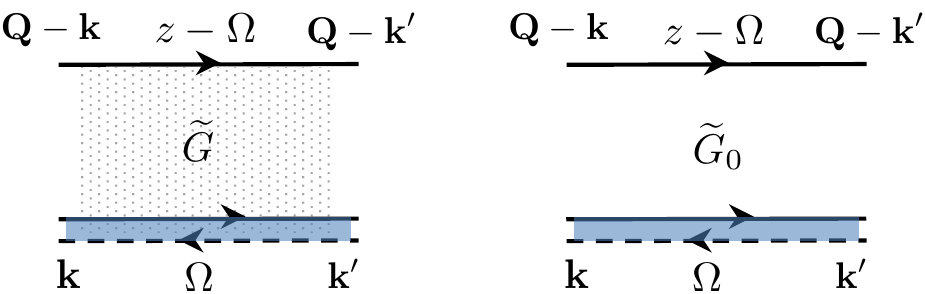}
\caption{Diagrammatic representation of the interacting (left) and noninteracting (right) Green's functions. The top and bottom propagators denote the Green’s functions of the electron and exciton, respectively.} \label{fig:g0g} 
\end{figure}

We use finite-temperature Green’s functions formalism to derive the BSE \cite{Mahan_Book,Haug_SchmittRink_PQE84}. Figure~\ref{fig:g0g} shows diagram representations of the interacting (left) and noninteracting (right) Green's functions of the exciton-electron system, $\widetilde{G}(\mathbf{Q},z;\mathbf{k}, \mathbf{k}', \Omega)$ and $\widetilde{G}_0(\mathbf{Q},z;\mathbf{k}, \mathbf{k}', \Omega)$. The total translational wavevector of the system is $\mathbf{Q}$. The exciton is a composite-boson and the electron is a fermion, and hence its total energy is given by a fermionic (odd) imaginary Matsubara energy, $z = (2\ell + 1)\pi i\beta^{-1}$, where $\ell$ is an integer and $\beta^{-1} =k_BT$ is the thermal energy. The double lines in the bottom of the diagrams in Fig.~\ref{fig:g0g} are the exciton propagators with electron and hole components. The incoming and outgoing translational wavevectors of the exciton are  $\mathbf{k}$ and $\mathbf{k}'$, respectively, and its energy is represented by a bosonic (even) imaginary Matsubara energy,  $\Omega = 2\ell\pi i\beta^{-1}$. The top lines in the diagrams of Fig.~\ref{fig:g0g} are the electron propagators where the incoming and outgoing wavevectors, $\mathbf{Q}-\mathbf{k}$ and $\mathbf{Q}-\mathbf{k}'$, as well as energy, $\Omega-z$, are such that the total translational wavevector and energy are conserved at $\mathbf{Q}$ and $z$.

To find the interacting Green's function of the exciton-electron system, we make use of the so-called ladder approximation which describes repeated interaction of the exciton with the electron through the matrix elements derived in the previous Appendix. The screened ladder approximation is used to explain experiments in which the semiconductor is neither subjected to intense photoexcitation nor to a large density of electrons. In other words, exciton condensation is irrelevant and the density of electrons is not large enough to dissociate trions through screening. Using the ladder approximation, the BSE is formally written as \cite{Haug_SchmittRink_PQE84,VanTuan_PRB24}
\begin{widetext}
 \begin{eqnarray}
\widetilde{G}(\mathbf{Q},z;\mathbf{k}, \mathbf{k}', \Omega)  &=&  \widetilde{G}_0(\mathbf{Q},z;\mathbf{k}, \mathbf{k}', \Omega) - \beta^{-1} \sum_{\mathbf{k}'', \mathbf{p},\Omega'} \widetilde{G}_0(\mathbf{Q},z;\mathbf{k}, \mathbf{k}'', \Omega) M(\mathbf{Q}, \mathbf{k}'', \mathbf{p}; \Omega-\Omega') \widetilde{G}(\mathbf{Q},z;\mathbf{p}, \mathbf{k}', \Omega').
 \label{eq:BSE_tilde}
\end{eqnarray}
\end{widetext}
Fortunately, the Green's functions can be contracted in two successive steps, rendering the problem tenable from a computation stand point. In the first step, we contract the function by integration out the outgoing wavevector ($\mathbf{k}'$). By defining 
 \begin{eqnarray}
\overline{G}(\mathbf{Q},z;\mathbf{k}, \Omega)  &=& \sum_{\mathbf{k}'} \widetilde{G}(\mathbf{Q},z;\mathbf{k},\mathbf{k}', \Omega),  \label{eq:contract1}
\end{eqnarray}
and using the fact that incoming and outgoing wavevectors are the same in a noninteracting Green's function, 
 \begin{eqnarray}
\widetilde{G}_0(\mathbf{Q},z;\mathbf{k}, \mathbf{k}', \Omega) &=& \overline{G}_0(\mathbf{Q},z;\mathbf{k}, \Omega) \delta_{\mathbf{k},\mathbf{k}'}  ,
 \label{eq:identity}
\end{eqnarray}
we arrive at
 \begin{eqnarray}
&& \!\!\!\!\!\!\!\! \overline{G}(\mathbf{Q},z;\mathbf{k}, \Omega)  =  \overline{G}_0(\mathbf{Q},z;\mathbf{k},  \Omega)  \nonumber \\ && \!\!\!\!\!\!\!\! \times \Big[ 1 -   \sum_{\mathbf{p},\Omega'} M(\mathbf{Q}, \mathbf{k}, \mathbf{p}; \Omega-\Omega') \overline{G}(\mathbf{Q},z;\mathbf{p}, \Omega') \Big]. 
 \label{eq:BSE_bar}
\end{eqnarray}
Next, we assume frequency-independent interaction $M(\mathbf{Q}, \mathbf{k}, \mathbf{p})=M(\mathbf{Q}, \mathbf{k}, \mathbf{p}; \Omega-\Omega')$, justified by the assumption that dynamical effects are irrelevant in the absence of other excitons or electrons in the system. We can then contract the Green's function for the second time by integrating out the bosonic degrees of freedom and analytically continuing  the conserved energy from the imaginary to real axis ($z \rightarrow E+i\delta$)
 \begin{eqnarray}
G(\mathbf{Q},E;\mathbf{k})  &=& -\beta^{-1}\sum_{\Omega} \overline{G}(\mathbf{Q},E;\mathbf{k},\Omega),  \label{eq:contract2}
\end{eqnarray}
getting that
 \begin{eqnarray}
&& G(\mathbf{Q},E;\mathbf{k}) =   G_0(\mathbf{Q},E;\mathbf{k})   \nonumber \\ &&\,\,\,+  \sum_{\mathbf{p}} G_0(\mathbf{Q},E;\mathbf{k}) M(\mathbf{Q}, \mathbf{k}, \mathbf{p}) G(\mathbf{Q},E;\mathbf{p}). 
\end{eqnarray}
Equation~(\ref{eq:BSEid}) of the paper reads the same, where the dependence of the Green's function on the conserved quantities $\mathbf{Q}$ and $E$ was implied in the main text. The last part is to find $G_0(\mathbf{Q},E;\mathbf{k})$ from contraction of $\overline{G}_0(\mathbf{Q},z;\mathbf{k},  \Omega)$. The noninteracting Green's function is the product of the exciton and electron Green's functions, 
 \begin{eqnarray}
{G}_0(\mathbf{Q},z;\mathbf{k}) &=& -\beta^{-1} \sum_{\Omega} \overline{G}_0(\mathbf{Q},z;\mathbf{k},\Omega) \nonumber \\ &=& -\beta^{-1} \sum_{\Omega}  G_e(\mathbf{Q}-\mathbf{k},z-\Omega) G_x(\mathbf{k},\Omega) . \qquad \,\, \label{eq:contract20}
\end{eqnarray}
$G_x(\mathbf{k},\Omega)$ and $G_e(\mathbf{Q}-\mathbf{k},z-\Omega)$ are the bottom and top propagators in Fig.~\ref{fig:g0g}. Writing this expression explicitly, we get 
\begin{widetext}
\begin{eqnarray}
{G}_0(\mathbf{Q},z;\mathbf{k})   &=&  -  \frac{1}{\beta} \sum_\Omega   {\color{black} \frac{1}{z- \Omega - \frac{\hbar^2 (\mathbf{Q}-\mathbf{k})^2}{2m'_{e}}  + \mu_\text{e} } } \,\,\,  {\color{black}  \frac{1}{\Omega - E_{_\text{X}} - \frac{\hbar^2 k^2}{2m_x}   + \mu_{_\text{X}}}   }   \nonumber \\
&=& \frac{1}{z  {\color{black} -  \frac{\hbar^2 (\mathbf{Q}-\mathbf{k})^2}{2m'_{e}} } {\color{black} - \frac{\hbar^2 k^2}{2m_x}} + \mu_0 }  \left( - \frac{1}{\beta} \sum_\Omega   {\color{black} \frac{1}{z- \Omega -  \varepsilon_{\mathbf{Q}-\mathbf{k}} } } \,\,\, -   \,\,\, \frac{1}{\beta} \sum_\Omega  {\color{black}  \frac{1}{\Omega - \hbar \omega_{\mathbf{k}}}}       \right)  \,\, \nonumber \\
&=& \frac{1 - f(\varepsilon_{\mathbf{Q}-\mathbf{k}}) - g(\hbar \omega_{\mathbf{k}})}{z  {\color{black} -  \frac{\hbar^2 (\mathbf{Q}-\mathbf{k})^2}{2m'_e} } {\color{black} - \frac{\hbar^2 k^2}{2m_x}} + \mu_0 }  . \,\, 
\end{eqnarray}
\end{widetext}
$\mu_{_\text{X}}$ and $\mu_{_\text{e}}$ are chemical potentials of the exciton and electron, respectively, and $E_{_\text{X}}$ is the binding energy of the exciton. In the transition between the first and second lines we have defined the energy $\mu_0 =  \mu_\text{e} + \mu_{_\text{X}} - E_{_\text{X}}$, and replaced the electron and exciton energies with $ \varepsilon_{\mathbf{Q}-\mathbf{k}}=  \hbar^2 (\mathbf{Q}-\mathbf{k})^2/2m'_{e} -  \mu_\text{e}  $ and $\hbar \omega_{\mathbf{k}}  =  E_{_\text{X}} + \hbar^2 k^2/2m_x - \mu_{_\text{X}}$.  In the transition between the second and last line, we have made use of the residue theorem to sum over Matsubara energies \cite{Mahan_Book}, where $f(\varepsilon_{\mathbf{Q}-\mathbf{k}})$ is the Fermi-Dirac distribution of electrons and $g(\hbar \omega_{\mathbf{k}})$ is the Bose-Einstein distribution of excitons. Taking $\mu_0=0$ to be the energy reference level, analytically continue $z$ to the real energy axis, and recalling that we are working in the dilute limit of electrons and excitons, $f(\varepsilon_{\mathbf{Q} -\mathbf{k}}) \rightarrow 0$ and $g(\hbar \omega_{\mathbf{k}} \rightarrow 0$), we arrive at Eq.~(\ref{eq:G0id}) of the main text.
\begin{eqnarray}
{G}_0(\mathbf{k})  \equiv {G}_0(\mathbf{Q},E;\mathbf{k})   &=& \frac{1}{ E -  \frac{\hbar^2 (\mathbf{Q}-\mathbf{k})^2}{2m'_e}  - \frac{\hbar^2 k^2}{2m_x} +  i\delta }   \,\,. \qquad
\end{eqnarray}
%%%%%%%%%%%%%%%%%%%%%%%%%%%%%%%%%%%%%%%%%%%%%%%%%%%%%%%%%%%%%%%%%%%%%%%%%%%%%%%
%%%%%%%%%%%%%%%%%%%%%%%%%%%%%%%%%%%%%%%%%%%%%%%%%%%%%%%%%%%%%%%%%%%%%%%%%%%%%%%
%%%%%%%%%%%%%%%%%%%%%%%%%%%%%%%%%%%%%%%%%%%%%%%%%%%%%%%%%%%%%%%%%%%%%%%%%%%%%%%
%%%%%%%%%%%%%%%%%%%%%%%%%%%%%%%%%%%%%%%%%%%%%%%%%%%%%%%%%%%%%%%%%%%%%%%%%%%%%%%

\section{Numerical details} \label{app:num}
As mentioned in the main text, the interacting Green's functions are solved for each energy value $E$ when $Q=0$. The latter renders the Green's functions isotropic and the computational effort is simplified. Equations~(\ref{eq:BSEid}), (\ref{eq:BSEdir}) and  (\ref{eq:BSEexc}) are solved through matrix inversion by recasting each of them to 
\begin{eqnarray}
\sum_{p}  \left[ \delta_{k,p} -   \,  \frac{AV_{0,k}}{4\pi^2} p\,dp \int_0^{2\pi} \!\!\!d \theta F({\bf k},{\bf p}) \right] V_p = V_{0,k}\,. \quad \label{eq:mat}
\end{eqnarray}
$\theta$ is the angle between ${\bf k}$ and ${\bf p}$, and $dp = p_{\text{max}}/N$ is the resolution of a uniform grid with $N$ points and cutoff wavevector $p_{\text{max}}$. Table \ref{tab} shows the assignments of $V_p$,  $V_{0,k}$ and $F({\bf k},{\bf p})$ based on the Equation we are solving, where $\delta_d=0$ is used in Eq.~(\ref{eq:BSEid}) and $\delta_d=1$ in Eq~(\ref{eq:BSEexc}). Note that the area $A$ in Eq.~(\ref{eq:mat}) is canceled out since the direct and exchange interactions are inversely proportional to the area.

\begin{table}[h]
\caption{\label{tab:BGREps} Assignments in Eq.~(\ref{eq:mat}).} \label{tab}
\begin{center}
\begin{tabular}{||c || c | c | c ||} 
 \hline
 Equation & $V_p$  & $V_{0,k}$ & $F({\bf k},{\bf p})$ \\ [0.5ex] 
 \hline\hline
(\ref{eq:BSEid})   & $G(p)$  & $G_0(k)$  & $D({\bf k},{\bf p}) + X(Q=0,{\bf k},{\bf p})$  \\
 \hline
\multirow{2}{*}{(\ref{eq:BSEdir})} & $G(p)$ & $G_0(k)$  & $D({\bf k},{\bf p})$  \\   & $G'(p)$  & $G_0'(k)$ & $D'({\bf k},{\bf p})$ \\
 \hline
\multirow{2}{*}{(\ref{eq:BSEexc})} & $g(p)$ & $G(k)$ & $\lambda(Q=0,{\bf k},{\bf p})$   \\  & $g'(p)$  & $G'(k)$  & $\lambda'(Q=0,{\bf k},{\bf p})$ \\
 \hline

 \hline
\end{tabular}
\end{center}
\end{table}

The number of grid points used in the simulations of Figs.~\ref{fig:res0}, \ref{fig:res28} and \ref{fig:id} is $N=1000$ and the wavevector cutoff is $k_{\text{max}}= p_{\text{max}} = 1/a_0$ when the ideal 2D Coulomb potential is used or $k_{\text{max}}= p_{\text{max}} = 0.1/a_0$ when the RK or $3\chi$ potentials are used ($a_0 = 0.53 \AA$). In addition,  the interacting Green's functions are solved with different energy values $E$ at a resolution of 0.1~meV. The solutions of Eqs.~(\ref{eq:BSEid}) and (\ref{eq:BSEdir}), shown in Fig.~\ref{fig:id} or by the dashed lines in Figs.~\ref{fig:res0} and \ref{fig:res28}, take about a minute with a simple laptop computer. The solution of Eq.~(\ref{eq:BSEexc}) takes several hours when $N=1000$ because of the tripled integration needed to evaluate each of the $N^2$ matrix elements in Eq.~(\ref{eq:lam}. For example,
\begin{eqnarray}
&&\!\!\!\!\!\!\!\!\! \int_0^{2\pi} \!\!\!d \theta  \lambda(Q=0,\mathbf{k},\mathbf{p})  =  \frac{A}{4\pi^2} \int_0^{2\pi} \!\! d \theta \int_0^{q_{\text{max}}} \!\!\!\! dq \,G'(q) \, q   \nonumber \\ 
&&\!\!\!\!\!\!\!\!\!  \times \int_0^{2\pi} \!\!\! d \varphi \, X(Q=0,k,q,\varphi) X'(Q=0,q,p,\varphi-\theta), \quad
 \label{eq:lam_term}
\end{eqnarray}
where $\varphi$ is the angle between $\mathbf{q}$ and $\mathbf{k}$, and $\varphi-\theta$ is the angle between $\mathbf{q}$ and $\mathbf{p}$.  A similar treatment is applied for the integrations of $\lambda'(Q=0,\mathbf{k},\mathbf{p})$, where Eq.~(\ref{eq:lam_term})  is written with  $X \leftrightarrow X'$ and $G(q)$ instead of $G'(q)$. %. \sum_{\mathbf{q}} X(\mathbf{Q},\mathbf{q},\mathbf{k})G'(\mathbf{q})X'(\mathbf{Q},\mathbf{p},\mathbf{q}), \nonumber \\
%\lambda'(\mathbf{Q},\mathbf{k},\mathbf{p})  &=&   \sum_{\mathbf{q}} X'(\mathbf{Q},\mathbf{q},\mathbf{k})G(\mathbf{q})X(\mathbf{Q},\mathbf{p},\mathbf{q}) .

%%%%%%%%%%%%%%%%%%%%%%%%%%%%%%%%%%%%%%%%%%%%%%%%%%%%%%%%%%%%%%%%%%%%%%%%%%%%%%%
%%%%%%%%%%%%%%%%%%%%%%%%%%%%%%%%%%%%%%%%%%%%%%%%%%%%%%%%%%%%%%%%%%%%%%%%%%%%%%%
%%%%%%%%%%%%%%%%%%%%%%%%%%%%%%%%%%%%%%%%%%%%%%%%%%%%%%%%%%%%%%%%%%%%%%%%%%%%%%%
%%%%%%%%%%%%%%%%%%%%%%%%%%%%%%%%%%%%%%%%%%%%%%%%%%%%%%%%%%%%%%%%%%%%%%%%%%%%%%%

\end{document}